\documentclass[final]{aa}

\usepackage{graphicx}
\usepackage{txfonts}
\usepackage{subcaption}         
\usepackage{lscape}             
\usepackage[colorlinks=true,
            linkcolor=blue,
            citecolor=blue,
            urlcolor=blue]{hyperref}
\usepackage[normalem]{ulem}
\def\INSPIRE{\mbox{{\tt INSPIRE}}}
\def\EINSPIRE{\mbox{{\tt E-INSPIRE}}}
\newcommand{\Msun}{\,$\mathrm{M}_\odot$}
\newcommand{\ppxf}{\textsc{pPXF}}
\defcitealias{PeraltaDeArriba+14}{PdA14}
\defcitealias{PeraltaDeArriba+15}{PdA15}

\begin{document}

\title{Beyond compactness:}
\subtitle{A structural--dynamical--evolutionary manifold \\ for the stellar-to-dynamical mass ratio in ultra-compact massive galaxies}

   \author{Chiara Spiniello\inst{1}\fnmsep\thanks{Corresponding author: chiara.spiniello@eso.org}}

   \institute{European Southern Observatory,  Karl-Schwarzschild-Stra\ss{}e 2, 85748, Garching, Germany}

   \date{Accepted 18 March 2026}

\abstract
{Ultra-compact massive galaxies (UCMGs) often exhibit elevated stellar-to-dynamical mass ratios when dynamical masses are estimated using standard virial prescriptions. This discrepancy has been interpreted as evidence for structural non-homology driven primarily by their compactness.}
{This study investigates how the stellar-to-dynamical mass ratio depends jointly on compactness ($\mathcal{C}$), internal kinematics ($\sigma_{\star}$), stellar population properties (mass-weighted age, metallicity, and [Mg/Fe]), and star formation histories. The analysis is based on a homogeneous catalogue of 482 UCMGs from the \INSPIRE\ and \EINSPIRE\ surveys, extending to significantly smaller sizes than previously analysed samples.}
{I first derive the compactness–mass relation assuming a constant virial coefficient ($K=5$). I then correct stellar masses for Initial Mass Function (IMF) variations and recompute stellar-to-dynamical mass ratios using an empirical prescription in which the virial coefficient varies as a function of radius and stellar mass. Finally, I test whether the relation is modulated by stellar kinematics and population properties, including the degree of relicness (DoR), which quantifies the extremeness of the star formation history.}
{A statistically significant anti-correlation between compactness and the IMF-corrected stellar-to-dynamical mass ratio is recovered when adopting a constant virial coefficient, even within the relatively narrow range of $\mathcal{C}$ spanned by nearby UCMGs. The relation substantially flattens when a structure-dependent $K$ is adopted, in agreement with previous literature. Beyond this one-dimensional behaviour, the data define a structural–dynamical manifold in the $(\log \mathcal{C}, \log \sigma_\star)$ space. Velocity dispersion sets the dominant axis of variation, and the corresponding plane accounts for $\sim62$\% of the variance in stellar-to-dynamical mass ratio. Including stellar age increases the explained variance to $\sim63$\%, revealing a secondary evolutionary modulation. In contrast, DoR, metallicity, and [Mg/Fe] do not retain independent explanatory power once stellar age is included.}
{The stellar-to-dynamical mass ratio in UCMGs is governed primarily by the depth of the gravitational potential, traced by stellar velocity dispersion, rather than by compactness alone. At fixed size, systems with higher $\sigma_\star$ exhibit systematically lower stellar-to-dynamical mass ratio, indicating that dynamical structure regulates the apparent mass imbalance in the ultra-compact regime. Compactness largely reflects this dynamical scaling, while stellar age introduces a coherent secondary modulation linking the structural manifold to the evolutionary state of the galaxy. Non-homology in UCMGs therefore encodes coupled dynamical and assembly processes rather than purely geometric compactness.}
\maketitle
\nolinenumbers

\section{Introduction}
In the local Universe, early-type galaxies (ETGs) follow a network of tight scaling relations linking stellar mass, effective radius, surface brightness, and velocity dispersion, including the Faber--Jackson relation \citep{FJ76}, the Fundamental Plane \citep{DD87, Dressler+87}, and the stellar mass--size relation \citep{Shen+03}. Modern dynamical studies have further reformulated these relations in terms of stellar and dynamical mass (viarial or model), giving rise to the so-called Mass Plane \citep{Cappellari+13_ATLAS3D_XV, MaNGADynPopIII_24}.

These scaling relations are commonly interpreted within the framework of virial equilibrium, where deviations from the simplest expectations are attributed to structural non-homology and variations in the stellar mass-to-light ratio \citep{Shen+03, Cappellari+06, Cappellari+13_ATLAS3D_XV}. In the simplest homologous approximation, galaxies are assumed to share similar internal density profiles and orbital structures, such that the virial coefficient $K$ entering dynamical mass estimates can be treated as approximately constant. 

For ETGs, detailed dynamical modelling has shown that a value $K \simeq 5$ provides an excellent approximation when velocity dispersion ($\sigma_{\star}$) is measured within one effective radius \citep{Cappellari+06}. In this regime, the virial estimate $M_{\rm dyn,vir}$ with $K=5$ closely reproduces the dynamical mass derived from full modelling, such that stellar ($M_\star$) and dynamical masses ($M_{\rm dyn}$) derived from full modelling are broadly consistent within observational uncertainties and systematic effects. However, analytical studies of Sérsic models have shown that the virial coefficient $K$ is an explicit function of the Sérsic index ($n$), reflecting variations in internal density structure and thereby breaking strict structural homology \citep{Ciotti91, BCD02}. Empirical and dynamical analyses confirm that fixed-$K$ estimators systematically deviate from more sophisticated dynamical masses as a function of $n$ \citep[e.g.][]{Zahid17,Frigo17,vanDerWel22}, and extended Fundamental Plane models account for structural non-homology traced by the Sérsic profile \citep[e.g.][]{Yehia25}. 
Systematic departures from simple virial expectations have been repeatedly reported in the literature, particularly among the most compact and massive systems. Several observational studies have shown that ultra-compact massive galaxies (UCMGs) can exhibit unusually high stellar-to-dynamical mass ratios, in some cases approaching or exceeding unity \citep{Martinez-Manso+11, PeraltaDeArriba+14, PeraltaDeArriba+15}. Such values are difficult to reconcile with standard assumptions about galaxy structure and mass distribution, and suggest that UCMGs may deviate systematically from the dynamical manifold defined by normal-sized ETGs. Recent IFU and dynamical studies further indicate that some ultra-compact systems can be strongly rotation-supported or otherwise non-spherical \citep[e.g.][]{Zhu25}, reinforcing the view that the virial estimator may be inadequate in such cases. 

A comprehensive investigation of this issue was presented by \citet{PeraltaDeArriba+14, PeraltaDeArriba+15}, hereafter PdA14 and PdA15  respectively, who analysed the dependence of the stellar-to-dynamical mass ratio on galaxy compactness across a heterogeneous sample of ETGs spanning a wide range of redshifts and structural properties. Compactness was quantified as the ratio between the observed effective radius and the size expected from the local stellar mass–size relation \citep{Shen+03}. They demonstrated that the apparent mass discrepancy correlates strongly with compactness, and showed that the relation can be largely reconciled by allowing the virial coefficient $K$ to vary systematically with compactness. In this picture, deviations from homology, represented by a non-constant $K$, become the primary driver of the observed discrepancy.

At the same time, increasing observational evidence indicates that UCMGs do not constitute a homogeneous population. For instance, the \INSPIRE\ \citep{Spiniello20_Pilot, Spiniello+21, DAgo23, Spiniello24} and \EINSPIRE\ \citep{Mills25} surveys have shown that galaxies with similar (large) stellar masses and (ultra-compact) sizes can display remarkably different stellar population properties, reflecting diverse formation histories (SFHs). While some compact systems show evidence of extended or rejuvenated star formation, others appear to have formed the bulk of their stellar mass at early cosmic epochs through an intense and rapid burst, subsequently evolving with minimal structural growth. These latter systems,  commonly referred to as relic galaxies \citep{Trujillo+09_superdense}, retain dense configurations, uniformly old and metal rich stellar populations, and enhanced [Mg/Fe] abundances \citep{Trujillo14, Ferre-Mateu+17, Spiniello+21, Spiniello24}. They have also been shown to have a larger stellar velocity dispersion than non-relics and normal-sized ETGs of similar stellar mass \citep{DAgo23, Spiniello24}. 

An open question is therefore whether the structural–dynamical relation identified by \citetalias{PeraltaDeArriba+14} and \citetalias{PeraltaDeArriba+15} applies uniformly across the full compactness spectrum, or whether it changes in the extreme regime occupied by relic galaxies. In particular, does the compactness–$M_{\star}/M_{\rm dyn}$ relation depend solely on structural parameters, as implied by non-homology arguments, or does it also reflect differences in stellar population properties and assembly history?

In this work, I reproduce the compactness and stellar-to-dynamical mass analysis of \citetalias{PeraltaDeArriba+14} and \citetalias{PeraltaDeArriba+15} using a homogeneous master catalogue of $\sim 500$ \INSPIRE\ and \EINSPIRE\ UCMGs. In addition to sizes and stellar masses, the dataset includes integrated stellar kinematics and detailed stellar population parameters (mass-weighted age and metallicity, and light-weighted SSP-like [Mg/Fe]), as well as the degree of relicness (DoR), a quantitative proxy for formation timescale. 
This ultra-compact and homogeneous sample allows, for the first time, a systematic assessment of whether deviations from strict homology in the extreme compact regime are driven purely by structural effects, as previously proposed, or whether they are additionally modulated by internal kinematics and assembly history.

This paper is organised as follows. Section~\ref{sec:data} describes the data, sample construction, and the derived physical parameters. In Section~\ref{sec:compactness-ratio}, I reproduce the compactness–dynamical mass relation and examine its residual dependence on stellar population and kinematic properties. Section~\ref{sec:hyperplane} introduces the structural–dynamical manifold in $(\log \mathcal{C}, \log \sigma_\star)$ space and quantifies its modulation by stellar age. Finally, Section~\ref{sec:discussion} discusses the physical implications of the results in the context of early galaxy assembly and relic evolution, and summarises the main conclusions.

\section{Data and sample}
\label{sec:data}
The analysis presented in this work is based on a homogeneous master catalogue constructed from the \INSPIRE\ and \EINSPIRE\ surveys. The first, based on an ESO Large Program (ID: 1104.B-0370, PI: C. Spiniello), was specifically designed to spectroscopically follow up and characterise the stellar populations of red UCMGs in the nearby Universe ($z<0.5$) identified via multi-band imaging \citep{Tortora+16_compacts_KiDS, Tortora+18_UCMGs, Scognamiglio20}. The latter extends the original programme by expanding the redshift coverage and wavelength range of \INSPIRE. Both datasets are publicly available on the author's webpage\footnote{\url{https://sites.google.com/inaf.it/chiara-spiniello/}}.

\begin{figure*}
\centering
\includegraphics[width=\textwidth]{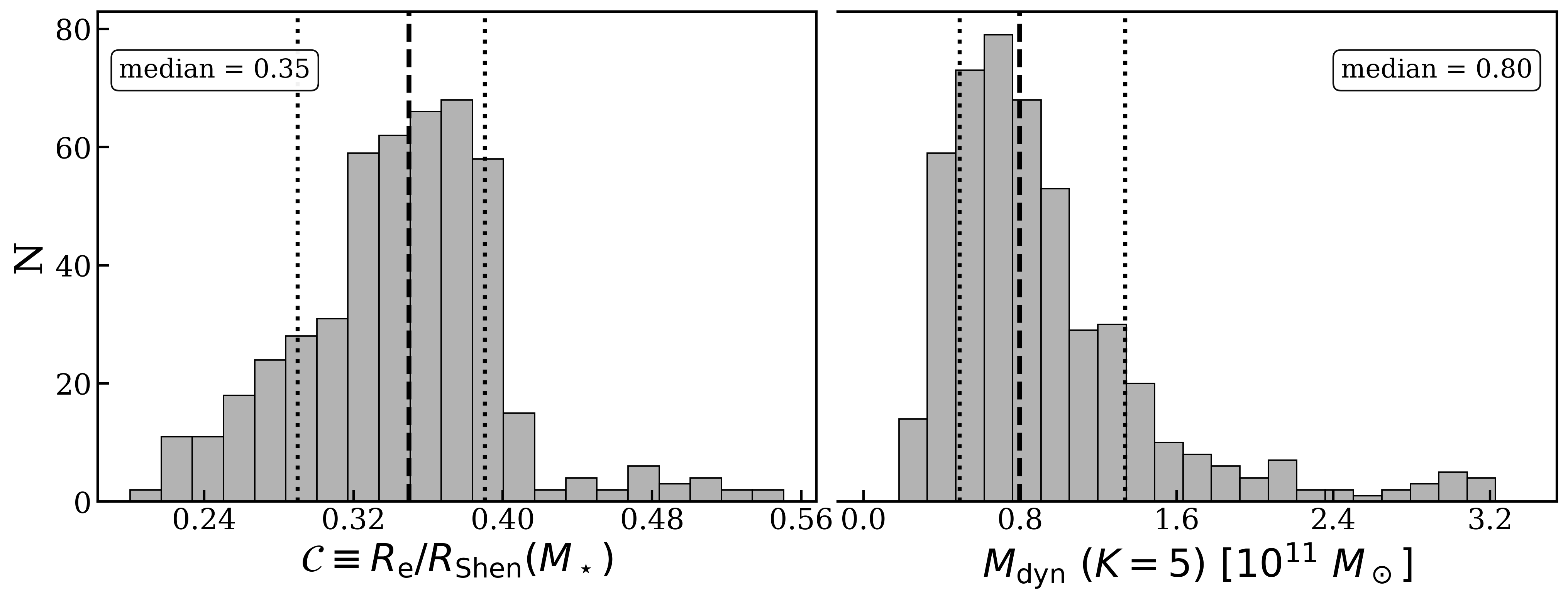}
\caption{
Distribution of compactness ($\mathcal{C} = R_{\rm e}/R_{\rm Shen}$; left panel) and virial dynamical mass ($M_{\rm dyn}$, $K=5$; right panel) for the full sample of 482 UCMGs. Vertical dashed lines indicate the median values, while dotted lines mark the 16th and 84th percentiles.
}
\label{fig:compactness_mdyn}
\end{figure*}

\subsection{The master catalogue of nearby UCMGs}
The catalogue combines all objects observed and analysed in the two surveys and presented in the most recent data releases \citep{Spiniello24, Mills25}, for which reliable measurements of structural parameters, stellar kinematics, and stellar population properties are available. By the time of writing, it represents the largest homogeneous sample of spectroscopically confirmed UCMGs with a measured DoR currently available in the nearby Universe.
The DoR is defined in \INSPIRE\ as a continuous, dimensionless parameter ranging from 0 to 1 that quantifies the assembly timescale of a galaxy’s stellar mass. It is derived from the reconstructed SFHs and combines the fraction of stellar mass formed within the first 3\,Gyr after the Big Bang with the cosmic times required to assemble 75\% and 99.8\% of the total stellar mass. High DoR values correspond to relics, i.e. systems that formed the bulk of their stellar mass rapidly at high redshift and subsequently experienced little or no additional star formation or structural growth, while lower DoR values indicate more extended or complex SFHs.

The sample comprises 482 UCMGs in the redshift range $0.03<z<0.41$, with effective radii ($R_{\rm e}$) smaller than 5 kpc and stellar masses larger than $3\times10^{10}$\,\Msun. It spans a broad range in DoR ($0.05<\mathrm{DoR}<0.89$), encompassing both relic and non-relic UCMGs. 
The galaxies cover mass-weighted ages between 2.3 and 12.4 Gyr, metallicities in the range $-0.32 < \mathrm{[M/H]} < +0.36$, and light-weighted SSP-like abundance ratios $0 < \mathrm{[Mg/Fe]} < 0.4$. Velocity dispersions are generally high, reaching $\sigma_{\star} > 500\,\mathrm{km\,s^{-1}}$ for the highest-DoR systems.

The 52 objects from \INSPIRE\ were selected using a fixed size threshold ($R_{\rm e} \le 2$\,kpc), while the majority of the sample, that comes from \EINSPIRE, satisfies the compactness criterion defined by \citet{Baldry21}, $\Sigma_{1.5} = M_{\star} R_{\rm e}^{-1.5}$. For these objects, effective radii derived from SDSS imaging were corrected to match those obtained from higher spatial resolution KiDS data, following the prescription in \citet{Mills25}. This ensures consistency across the combined dataset. 

The effective radii of all UCMGs in the current catalogue lie well below those predicted by the local stellar mass–size relation, placing the sample firmly in the compact tail of the ETG distribution (see, e.g., Figure~1 of \citealt{Mills25}). In contrast to the heterogeneous sample analysed by \citetalias{PeraltaDeArriba+14}, which is dominated by typical SDSS ETGs, sparsely samples the most compact systems, and extends to higher redshift, \INSPIRE\ and \EINSPIRE\ deliberately focus on the ultra-compact regime while restricting the analysis to the nearby Universe ($z<0.5$). 
Hence, the combination of extreme compactness, high velocity dispersions, and diverse stellar populations on a homogeneous sample of nearby  UCMGs, makes this dataset particularly well-suited to assess whether departures from strict homology are driven solely by structural effects or are additionally modulated by assembly history.

\subsection{Compactness and dynamical masses}
\label{sec:derived_quantities}
Following \citetalias{PeraltaDeArriba+14}, galaxy compactness is quantified here as the ratio between the observed effective radius and the expected size from the local stellar mass–size relation:

\begin{equation}
\mathcal{C} = \frac{R_{\rm e}}{R_{\rm Shen}(M_{\star})}.
\end{equation}

Here $R_{\rm Shen}(M_{\star})$ is the median size at fixed stellar mass given by \citet{Shen+03}, expressed for stellar masses computed assuming a Kroupa Initial Mass Function (IMF). Since the  stellar masses from the catalogue are also derived adopting a Kroupa IMF, no additional IMF conversion is applied\footnote{In \citetalias{PeraltaDeArriba+14}, the authors apply a correction to consider masses computed assuming a Salpeter-like IMF instead.}.
Although more recent mass–size calibrations exist \citep[e.g.][]{vanderWel14, Lange+15,Mowla19,Kawinwanichakij21}, I adopt \citet{Shen+03} for consistency with \citetalias{PeraltaDeArriba+14}.  
With the above definition, by construction, galaxies with $\mathcal{C}<1$ are more compact than the local early-type population at the same stellar mass.
In physical terms, $C$ acts as a proxy for central mass concentration relative to typical ETGs.

Throughout this work I use the virial dynamical mass estimator,
\begin{equation}
M_{\rm dyn,vir} = K\,\frac{\sigma_{\star}^{2} R_{\rm e}}{G},
\end{equation}
where $\sigma_{\star}$ is the stellar velocity dispersion, $R_{\rm e}$ the effective radius, $G$ the gravitational constant, and $K$ the virial coefficient. In the text and figures below, whenever we refer to ``dynamical mass'' computed from the data we explicitly mean this virial estimate, $M_{\rm dyn,vir}$. We do not attempt to recover the true dynamical mass $M_{\rm dyn}$ that would be obtainable from full dynamical modelling (e.g. JAM or Schwarzschild methods) because such methods require spatially resolved kinematic maps that are unavailable for the full UCMG sample. Differences between $M_{\rm dyn,vir}$ and $M_{\rm dyn}$ can arise from non-homology, anisotropy, rotation, or departures from spherical symmetry \citep[e.g.][]{Cappellari+06, Zhu25}, and these issues motivate the empirical and robustness tests presented in Appendix~\ref{app:empirical_fit}.

Integrated velocity dispersion values were measured from the \INSPIRE\ and \EINSPIRE\ spectra using full-spectrum fitting with the software \ppxf\ \citep{Cappellari17}, and then corrected for aperture effects following the prescription in \citet{Cappellari+06}. In detail, for \INSPIRE\ the R$_{50}$ apertures defined in the survey papers are adopted, while for SDSS spectra velocity dispersions are aperture-corrected assuming a fixed radius of 1.5$''$ (corresponding to half of the SDSS fibre diameter), after correcting the effective radii for the worse spatial resolution, as described in \citet{Mills25}. It is worth noting that in all cases the apertures are larger than the physical effective radii of the UCMGs. 
Uncertainties on virial dynamical masses were propagated from the measured uncertainties on velocity dispersion and assuming a conservative 15\% on the stellar masses and 25\% uncertainty on the effective radii to account for the difficulty of measuring such quantities from ground-based multi-band imaging. Further details on measurements and error estimation can be found in the \INSPIRE\ and \EINSPIRE\ survey papers, as well as in the Appendix of \citet{Tortora+16_compacts_KiDS}. To enable a direct comparison with \citetalias{PeraltaDeArriba+14} and \citetalias{PeraltaDeArriba+15}, I initially adopt their baseline assumption of a constant virial coefficient $K=5$, corresponding to the homologous approximation calibrated for local ETGs \citep{Cappellari+06}. However, the impact of adopting a structure-dependent virial coefficient, as done in  these papers, is explored in Section~\ref{sec:K_PdA}. 

The derived structural properties of the full sample are summarised in Figure~\ref{fig:compactness_mdyn}, which shows the distribution of compactness ($\mathcal{C} = R_{\rm e}/R_{\rm Shen}$) and dynamical masses. As anticipated, the sample is strongly skewed towards compact systems with $\mathcal{C}<0.5$, confirming that it predominantly probes the extreme tail of the local ETG size distribution. The median compactness is $\mathcal{C}\simeq0.35$, with only a small fraction of objects approaching $\mathcal{C}\sim0.6$. The corresponding dynamical mass distribution peaks around $M_{\rm dyn}\sim8\times10^{10}\,M_\odot$, extending up to $\sim3\times10^{11}\,M_\odot$ for the most massive systems. 

\section{The compactness–stellar-to-dynamical mass relation}
\label{sec:compactness-mass}
In this section, I examine how the stellar-to-dynamical mass ratio depends on compactness in the ultra-compact regime. \citetalias{PeraltaDeArriba+14} and \citetalias{PeraltaDeArriba+15} demonstrated that the apparent mass discrepancy correlates strongly with compactness and can be largely reconciled by adopting a structure-dependent virial coefficient. In their analysis, however, stellar masses were computed assuming a fixed IMF, such that deviations from homology were interpreted primarily in terms of dynamical structure.

Here, I extend this framework by additionally accounting for IMF variations that correlate with formation history in UCMGs. Both the stellar mass normalisation and the dynamical mass calibration may therefore influence the inferred compactness–$M_\star/M_{\rm dyn}$ relation.

\subsection{IMF variations and stellar mass corrections}
\label{sec:IMF}

Recent results from the \INSPIRE\ collaboration have demonstrated that the stellar IMF slope in UCMGs is not universal, but correlates with the degree of relicness (DoR) \citep{Martin-Navarro+23, Maksymowicz-Maciata24}. In particular, \citet{Maksymowicz-Maciata24} derived the empirical relation

\begin{equation}
\Gamma_b = 3.42\,\mathrm{DoR} + 0.15,
\end{equation}

where $\Gamma_b$ is the low-mass IMF slope. 

Adopting this calibration, variations in IMF slope are converted into stellar mass corrections via

\begin{equation}
\Delta \log M_\star = k_{\rm IMF} \, (\Gamma_b - \Gamma_{b,\mathrm{Kroupa}}),
\end{equation}

where $\Gamma_{b,\mathrm{Kroupa}} = 1.3$ corresponds to a Kroupa IMF. The coefficient $k_{\rm IMF}$ reflects the sensitivity of the stellar mass-to-light ratio to IMF slope variations. For old, metal-rich stellar populations typical of UCMGs, stellar population synthesis models based on the MILES library predict 
$\partial \log(M_\star/L)/\partial \Gamma_b \sim 0.15$--$0.20$ 
\citep[e.g.][]{Vazdekis+12, Vazdekis15, Conroy_vanDokkum12a}. 
A representative value $k_{\rm IMF}=0.18$ is adopted.

To avoid extrapolation beyond the calibrated regime and because no evidence is found for IMFs lighter than Kroupa in this class of systems, a conservative floor $\Delta \log M_\star \ge 0$ is imposed. Stellar masses are therefore not reduced below the Kroupa reference.

Uncertainties in the IMF–DoR calibration and in DoR measurements are propagated via Monte Carlo (MC) sampling; the implementation details are described in Appendix~\ref{app:IMF_MC}. For all subsequent analyses, the IMF-corrected stellar mass is taken as the median of the MC distribution, and stellar-to-dynamical mass ratios are computed consistently for each realization.

\begin{figure}
\centering
\includegraphics[width=\columnwidth]{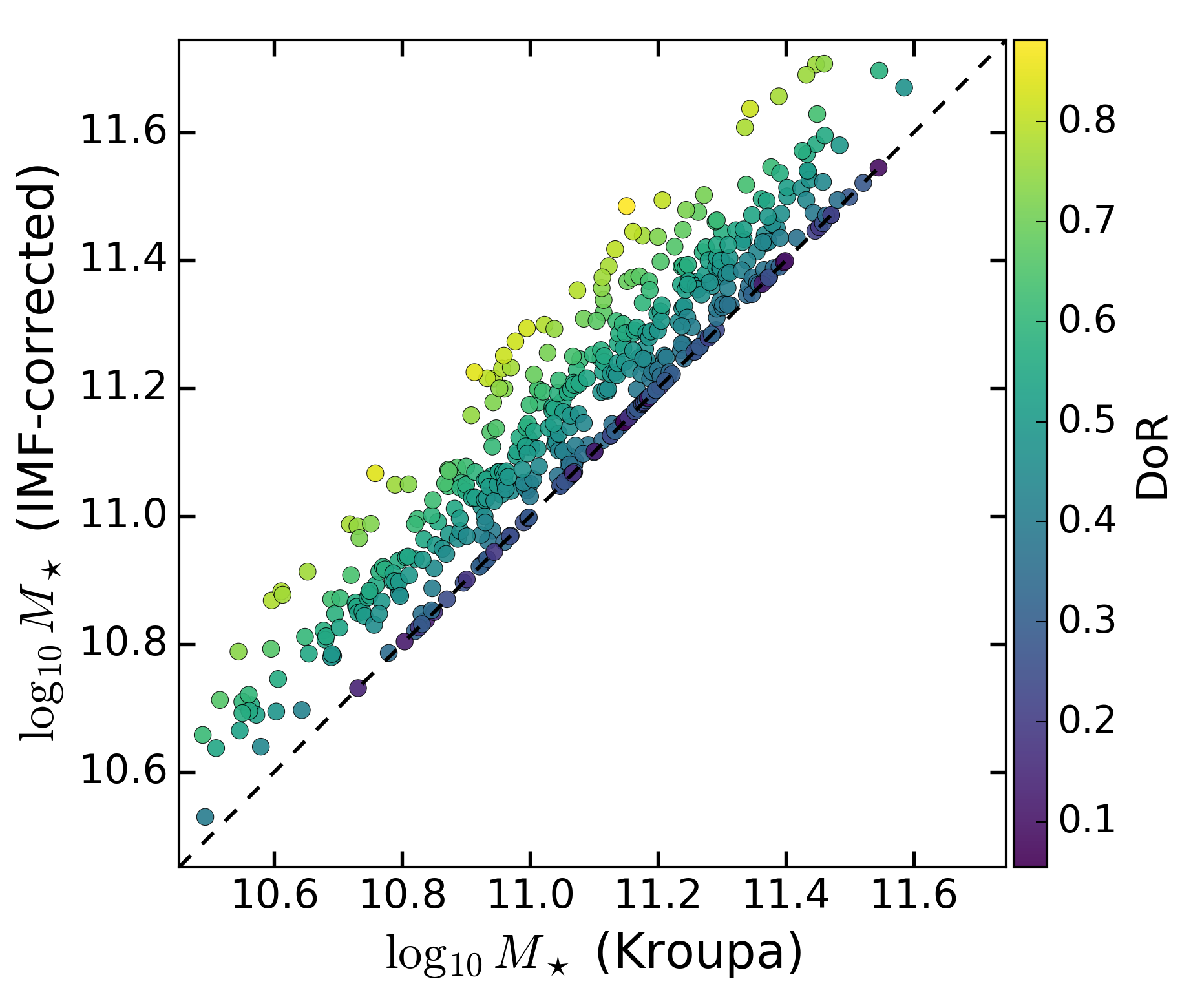}
\caption{
Comparison between stellar masses derived assuming a Kroupa IMF and those corrected according to the empirical IMF–DoR relation. Points are colour-coded by DoR and the dashed line indicates the one-to-one relation. 
}
\label{fig:imf_mass_shift}
\end{figure}

Figure~\ref{fig:imf_mass_shift} illustrates the object-by-object stellar-mass shift induced by the IMF correction. The adjustment reaches up to $\sim 0.3$ dex for the most extreme relic systems and is expected to influence the inferred $M_\star/M_{\rm dyn}$ ratios. The impact on the compactness--mass relation is examined in Section~\ref{sec:final_comp-mass}.

\subsection{Structure-dependent virial coefficient}
\label{sec:K_PdA}

The dynamical masses derived in Section~\ref{sec:derived_quantities} have been computed assuming a constant virial coefficient $\mathrm{K}=5$, following the standard homologous approximation calibrated for local ETGs \citep{Cappellari+06}. However, \citetalias{PeraltaDeArriba+14} showed that compact massive galaxies are expected to deviate from strict structural homology, and that a constant virial coefficient is therefore inadequate in this regime.

Hence, I next consider a structure-dependent virial coefficient that explicitly accounts for the dependence of the dynamical mass estimator on galaxy size and stellar mass. In this framework, the virial coefficient is parametrised as

\begin{equation}
\mathrm{K}_{\mathrm{PdA}} = 6.0
\left(\frac{R_{\rm e}}{3.185\,\mathrm{kpc}}\right)^{-0.81}
\left(\frac{M_\star}{10^{11}\,M_\odot}\right)^{0.45},
\end{equation}

following Eq.~11 in \citetalias{PeraltaDeArriba+14}.  

Dynamical masses are therefore recomputed as

\begin{equation}
M_{\rm dyn, PdA} = \mathrm{K}_{\mathrm{PdA}}
\frac{\sigma_{\star}^{2} R_{\rm e}}{G},
\end{equation}

allowing a direct comparison between the constant-$\mathrm{K}$ and structure-dependent-$\mathrm{K}$ scenarios.

It is important to note that in this subsection the PdA coefficient is evaluated using stellar masses computed under the baseline Kroupa IMF. 
This choice was made to isolate the effect of structural non-homology alone, without introducing additional IMF-driven shifts. The combined effect will then be evaluated in the next Section. 

\begin{figure}
\centering
\includegraphics[width=\columnwidth]{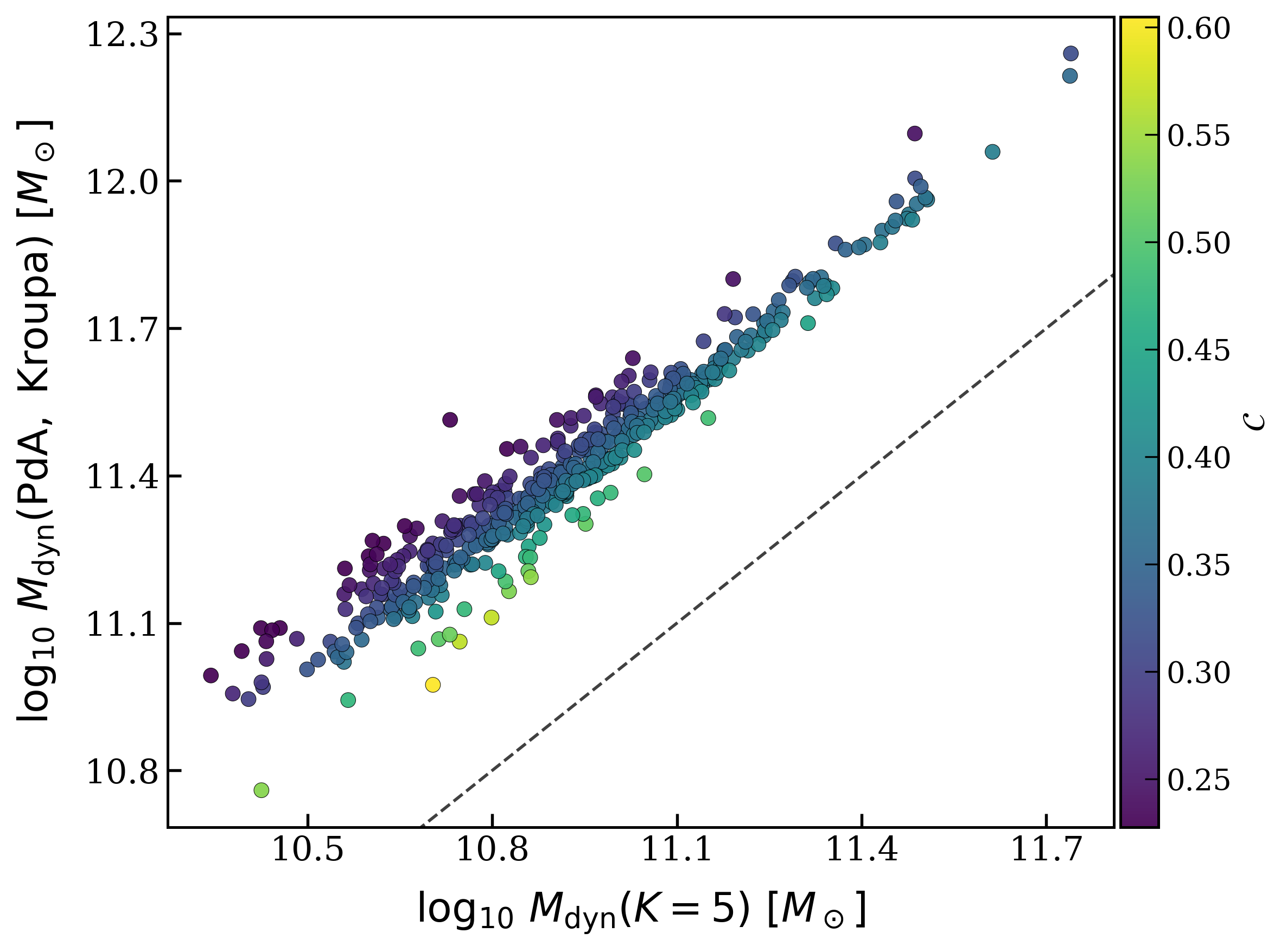}
\caption{
Comparison between dynamical masses computed assuming a constant virial coefficient ($K=5$) and those obtained using the structure-dependent calibration of \citetalias{PeraltaDeArriba+14}. The dashed line indicates the one-to-one relation. Points are colour-coded by compactness.}
\label{fig:masses_PdA}
\end{figure}

To visualise the magnitude of the dynamical mass shift induced by the structure-dependent virial calibration, Figure~\ref{fig:masses_PdA} directly compares dynamical masses computed under the homologous assumption ($K=5$) with those obtained using the PdA prescription. The dashed line indicates the one-to-one relation. Points are colour-coded by compactness.

All systems lie systematically above the one-to-one line, indicating that the structure-dependent calibration increases the inferred dynamical mass relative to the constant-$K$ estimate. The shift is monotonic and largest for the most compact systems, as reflected by the colour gradient. This behaviour demonstrates that deviations from structural homology introduce a substantial and compactness-dependent rescaling of the dynamical mass scale prior to analysing stellar-to-dynamical mass ratios. In the following subsection, I quantify how this structural rescaling propagates into the compactness--$M_\star/M_{\rm dyn}$ relation.

\subsection{Impact of virial calibration on the compactness–mass relation}
\label{sec:final_comp-mass}

\begin{figure}
\centering
\includegraphics[width=\linewidth]{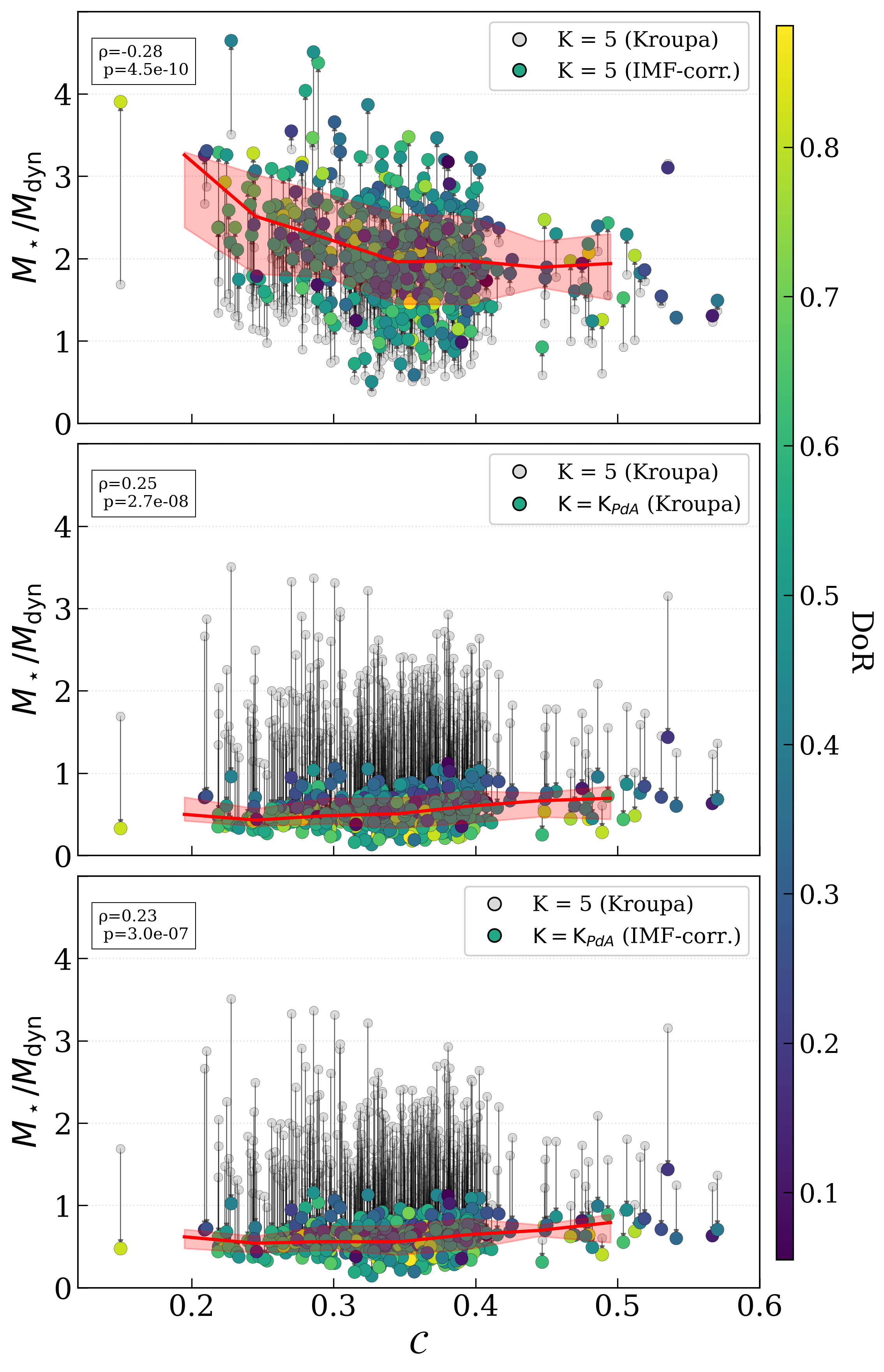}
\caption{
Stellar-to-dynamical mass ratio as a function of compactness under different assumptions. In all panels, points are colour-coded by DoR, while grey symbols indicate non-corrected masses, with arrows marking correction-driven shifts. 
Top: IMF-corrected stellar masses with a constant virial coefficient $K=5$. 
Middle: Structure-dependent virial coefficient $K=K_{\rm PdA}$ with Kroupa stellar masses. 
Bottom: Fully adjusted configuration combining $K=K_{\rm PdA}$ with IMF-corrected stellar masses.}
\label{fig:compactness}
\end{figure}

The consequences of adopting a structure-dependent virial coefficient and of correcting stellar mass estimates for a non-universal IMF slope, are illustrated in Figure~\ref{fig:compactness}, which presents a three-panel comparison of the stellar-to-dynamical mass ratio as a function of compactness.

The top panel shows the IMF-corrected stellar-to-dynamical mass ratios computed under the standard homologous assumption $\mathrm{K}=5$, colour-coded by DoR. Grey symbols represent the original Kroupa-based stellar masses, with arrows indicating the IMF-driven shifts. Although IMF-driven corrections increase $M_\star$ for high-DoR systems, a high fraction of galaxies still exhibit $M_\star/M_{\rm dyn} > 1$, particularly at low compactness values. This indicates that IMF variations alone are insufficient to reconcile the mass discrepancy in the ultra-compact regime.

The middle panel instead adopts the structure-dependent prescription $\mathrm{K}=\mathrm{K}_{\mathrm{PdA}}$, while retaining Kroupa stellar masses. Grey symbols again represent the uncorrected $K = 5$ ratios. Under this calibration, the previously unphysical regime $M_\star/M_{\rm dyn} > 1$ is substantially reduced. This demonstrates that deviations from structural homology, encapsulated in the virial coefficient, have a major impact on the inferred dynamical mass scale. The effect of the structural rescaling is significantly larger than that induced by IMF corrections alone.

In addition, a clear colour segregation is visible in this panel: at fixed compactness, high-DoR systems (yellow–green points) tend to exhibit systematically lower $M_\star/M_{\rm dyn}$ ratios when Kroupa stellar masses are adopted. This behaviour reflects the fact that relic-like systems are expected to host steeper IMFs, such that their stellar masses are underestimated under a universal Kroupa assumption. The structural correction alone therefore does not remove this population-dependent modulation.

The bottom panel combines $\mathrm{K}=\mathrm{K}_{\mathrm{PdA}}$ with IMF-corrected stellar masses. In this fully adjusted configuration, the stellar-to-dynamical mass ratios are largely confined to a physically plausible range, with the previously extreme $M_\star/M_{\rm dyn} > 1$ regime strongly suppressed, although a small number of objects still lie near or slightly above unity. The colour segregation observed in the middle panel is correspondingly reduced once IMF corrections are applied, confirming that IMF variations introduce an additional, physically motivated modulation of the mass ratio beyond structural non-homology. 
However, it is important to emphasise here that the PdA scaling is not derived from first-principles Jeans modelling, but represents an empirical calibration designed to restore consistency between stellar and dynamical mass estimates in the compact regime.

Taken together, these results indicate that the assumption of a constant virial coefficient $\mathrm{K}=5$, calibrated for local normal-sized homologous ETGs, is not appropriate for the ultra-compact systems probed here. Even after accounting for IMF-driven stellar mass variations, a structure-dependent virial calibration remains necessary to obtain physically consistent stellar-to-dynamical mass ratios for UCMGs.

Importantly, although adopting $\mathrm{K}=\mathrm{K}_{\rm PdA}$ substantially reduces the previously unphysical regime $M_{\star}/M_{\rm dyn}>1$, it also modifies both the sign and amplitude of the compactness trend. Under the homologous assumption ($K=5$), I recover a significant negative correlation between $M_{\star}/M_{\rm dyn}$ and $\mathcal{C}$, such that more compact systems display larger stellar-to-dynamical mass ratios, in qualitative agreement with previous literature. Once $\mathrm{K}_{\rm PdA}$ is adopted, this large-scale anti-correlation largely disappears: the best-fitting relation becomes much flatter and the Spearman rank coefficient becomes small and positive for both IMF-corrected and non-corrected stellar masses. This modest but non-zero residual indicates that structural non-homology explains the bulk of the discrepancy, while a weaker dependence on compactness, or on correlated structural or dynamical quantities, remains.

To assess whether this residual trend reflects formation history or stellar population diagnostics, the next section examines the residuals from the structure-dependent fit as a function of DoR, age, [M/H], [Mg/Fe], and velocity dispersion.

\section{Beyond compactness: stellar kinematic and population imprints}
\label{sec:compactness-ratio}

The previous analysis has shown that adopting a structure-dependent virial coefficient 
($\mathrm{K}_{\rm PdA}$) and correcting stellar masses for IMF variations 
largely removes the strong anti-correlation between $M_\star/M_{\rm dyn}$ and compactness 
that emerges under the homologous assumption ($K=5$). 
Structural non-homology therefore accounts for the bulk of the apparent mass discrepancy in UCMGs.

Nevertheless, a residual dispersion remains around the compactness–mass relation, 
indicating that size alone does not provide a complete description of the stellar-to-dynamical mass ratio. 
Galaxies with similar compactness may differ in their internal mass concentration, 
assembly epoch, stellar population properties, and kinematic structure. 
To assess whether such parameters introduce additional systematic trends, 
I examine the residuals of the compactness–mass relation after removing the dependence on $\mathcal{C}$.

The residuals are defined as deviations from a robust fit of

\begin{equation}
\log_{10}\!\left(\frac{M_\star}{M_{\rm dyn, PdA}}\right)
= a + b\,\log_{10}(\mathcal{C}),
\end{equation}

where the fit is computed using IMF-corrected stellar masses and 
structure-dependent dynamical masses ($\mathrm{K}=\mathrm{K}_{\rm PdA}$).
The residual for each galaxy is therefore defined as

\begin{equation}
\Delta \log_{10}\!\left(\frac{M_{\star,\mathrm{IMF}}}{M_{\rm dyn,PdA}}\right)
=
\log_{10}\!\left(\frac{M_{\star,\mathrm{IMF}}}{M_{\rm dyn,PdA}}\right)
-
\left[a+b\,\log_{10}(C)\right]
\end{equation}

and quantifies the departure of each system from the mass ratio predicted solely on the basis of its compactness.

\subsection{Formation history and stellar population imprint}

\begin{figure*}
\centering
\includegraphics[width=\linewidth]{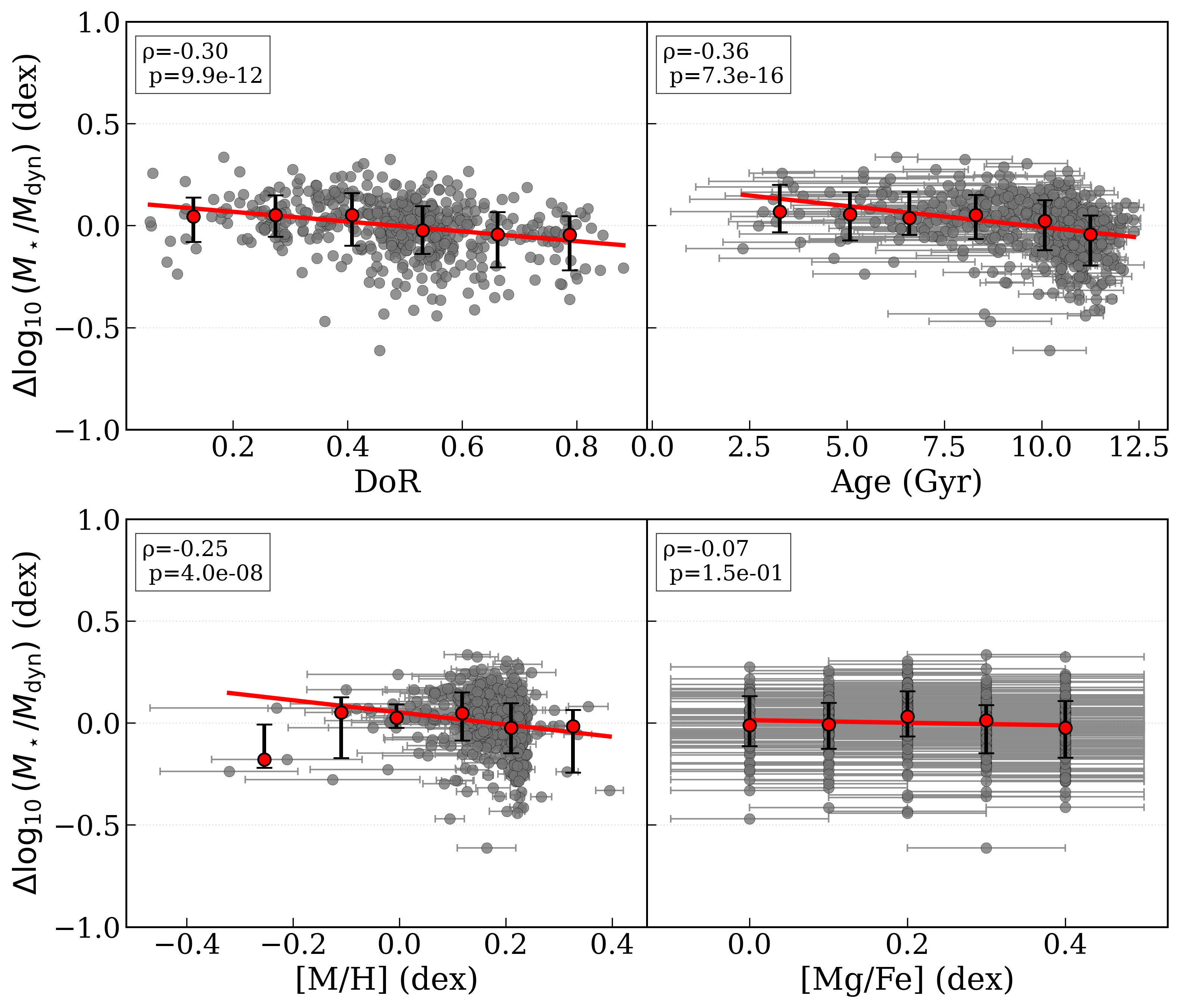}
\caption{
Residuals of the compactness--mass relation,
$\Delta \log_{10}(M_\star/M_{\rm dyn})$, 
as a function of stellar population and formation-history indicators:
DoR (top left), mass-weighted age (top right),
metallicity [M/H] (bottom left), and SSP-like [Mg/Fe] (bottom right).
Grey points represent individual galaxies.
Red symbols indicate binned medians with 16--84\% intervals,
and red lines show robust linear fits.
Spearman coefficients are reported in each panel.
}
\label{fig:residuals_stelpop}
\end{figure*}

Figure~\ref{fig:residuals_stelpop} explores whether stellar population properties 
introduce systematic modulation of the structural relation. 
Moderate but statistically significant trends are observed with both 
DoR and mass-weighted stellar age. 
At fixed compactness, galaxies that assembled their stellar mass 
earlier and more rapidly tend to exhibit systematically lower 
$M_\star/M_{\rm dyn}$ ratios than predicted by the structural fit. 
The residuals therefore retain an imprint of formation epoch.

No statistically significant dependence is detected with [Mg/Fe], which traces the duration of star formation episodes \citep{Thomas+05}.  
Since mass-weighted age and DoR encode instead the global epoch of assembly,  the residual behaviour appears more closely linked 
to the timing of stellar mass build-up 
than to enrichment timescales alone. 
It should nevertheless be noted that the larger measurement uncertainties in [Mg/Fe] reduce the effective dynamic range of this diagnostic, which may partly contribute to the absence of a statistically significant correlation.

A weaker but still statistically significant correlation 
is found with metallicity: more metal-rich systems 
tend to lie slightly below the compactness-predicted relation. 
However, the [M/H] distribution is relatively concentrated in this sample, with the most extreme bins containing a limited number of objects. The inferred metallicity trend should therefore also be interpreted with caution.

Overall, stellar population parameters introduce a measurable but secondary modulation of the compactness–mass relation. 
Even after accounting for structural non-homology and IMF variations, 
galaxies that formed earlier retain a systematically different 
stellar-to-dynamical mass balance. 
These effects, however, remain modest in amplitude relative to the intrinsic scatter of the relation.

\subsection{Kinematic structure as the dominant residual axis}
\label{sec:sigma}

\begin{figure}
\centering
\includegraphics[width=\columnwidth]{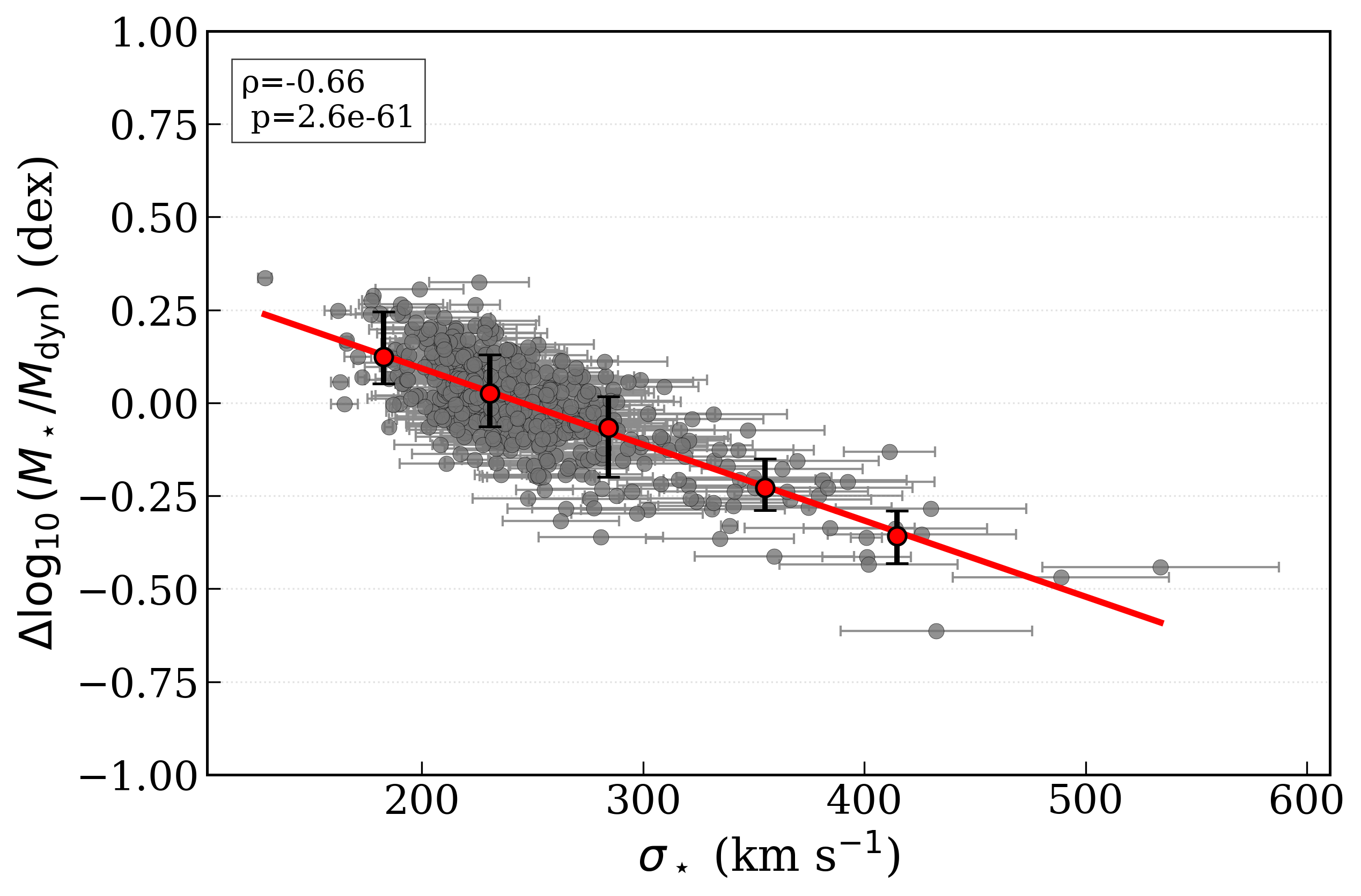}
\caption{
Residuals of the compactness--mass relation 
as a function of central velocity dispersion $\sigma_\star$.
Symbols and annotations follow the conventions of Fig.~\ref{fig:residuals_stelpop}.
}
\label{fig:residuals_sigma}
\end{figure}

In contrast to the moderate population trends, the most pronounced residual dependence emerges with velocity dispersion, as shown in Figure~\ref{fig:residuals_sigma}.  
A strong and highly significant anti-correlation ($\rho=-0.66$) is observed: 
galaxies with larger $\sigma_\star$ systematically lie below 
the compactness-predicted $M_\star/M_{\rm dyn}$ relation.

Although $M_{\rm dyn}$ formally depends on $\sigma_\star^2$, 
this behaviour is not a trivial algebraic consequence. 
Compactness already encapsulates deviations in size at fixed stellar mass, 
and the structure-dependent virial calibration explicitly accounts 
for systematic size- and mass-scaling. 
The persistence of a strong residual–$\sigma_\star$ correlation 
therefore indicates a genuine diversity in internal dynamical structure 
among galaxies with similar compactness.

Physically, $\sigma_\star$ traces the depth of the gravitational potential 
and is directly related to the central mass distribution \citep{Cappellari16}. 
Variations in mass concentration, inner density slope, 
orbital anisotropy, or dark-matter fraction within the effective radius 
can modify the mapping between the observed line-of-sight dispersion 
and the total kinetic energy. 
Systems with higher velocity dispersions may therefore possess more centrally concentrated mass configurations 
than captured by the simple virial scaling, leading to systematically lower inferred $M_\star/M_{\rm dyn}$ at fixed compactness.

Importantly, stellar masses have already been corrected for IMF variations (Sec.~\ref{sec:IMF}). The strong residual–$\sigma_\star$ trend cannot therefore be attributed to IMF-driven mass shifts. 
Instead, it reveals that internal kinematics encodes structural information that is not captured by compactness alone. 
The strength and structural nature of the $\sigma_\star$ dependence indicate that internal kinematics must be treated as a primary coordinate in the mass–discrepancy relation. 
Compactness by itself does not define the full structural scaling of $M_\star/M_{\rm dyn}$ in UCMGs, motivating a multivariate description in which size and internal kinematics are considered simultaneously.

\section{A structural manifold in $(\log \mathcal{C}, \log \sigma_\star)$ space}
\label{sec:hyperplane}

The strong residual–$\sigma_\star$ correlation identified in Sect.~\ref{sec:sigma} demonstrates that compactness alone does not provide a complete description of the stellar-to-dynamical mass ratio. I therefore quantify the joint dependence on size and internal kinematics through a multivariate regression of the form

\begin{equation}
\log_{10}\left(\frac{M_{\star,\mathrm{IMF}}}{M_{\rm dyn,PdA}}\right)
=
a
+
b\,\log_{10}\mathcal{C}
+
c\,\log_{10}\sigma_{\star \mathrm{,corr}},
\label{equation}
\end{equation}

where $\sigma_{\star \mathrm{,corr}}$ is the aperture-corrected velocity dispersion. All coefficients are estimated using a robust M-estimator with Huber loss \citep{Huber1964,HuberRonchetti2009}. Explained variance is computed from the predictions of the robust fit as

\begin{equation}
R^2_{\rm RLM} = 1 - \frac{\mathrm{Var}(y-\hat{y}_{\rm RLM})}{\mathrm{Var}(y)} ,
\end{equation}

where Var denotes the population variance (ddof = 0), ensuring consistency with the bootstrap computation. Parameter uncertainties are estimated via bootstrap resampling ($n=1000$).

\begin{table*}
\caption{Robust multivariate regressions for $\log(M_{\star,\mathrm{IMF}}/M_{\rm dyn,PdA})$.}
\label{tab:regressions}
\centering
\begin{tabular}{lccccccc}
\hline\hline
Model & const & $b$ (log $\mathcal{C}$) & $c$ (log $\sigma_{\star}$) & $d_{\rm Age}$ & $d_{\rm DoR}$ & $R^2_{\rm RLM}$ & RMS$_{\rm RLM}$ \\
\hline
(1) Compactness only 
& $-0.040$ 
& $0.483$ 
& -- 
& -- 
& -- 
& 0.0547 
& 0.1380 \\

(2) Structural plane 
& $3.149$ 
& $0.173$ 
& $-1.389$ 
& -- 
& -- 
& 0.6170 
& 0.0877 \\

(3) + Age 
& $3.096$ 
& $0.142$ 
& $-1.332$ 
& $-0.0099$ 
& -- 
& 0.6317 
& 0.0860 \\

(4) + DoR 
& $3.172$ 
& $0.176$ 
& $-1.401$ 
& -- 
& $0.016$ 
& 0.6173 
& 0.0876 \\
\hline
\end{tabular}
\end{table*}

Results of the regression are summarised in Table~\ref{tab:regressions}, where variance and RMS are computed from RLM predictions. Uncertainties correspond to 95\% bootstrap confidence intervals ($n=1000$). 
The compactness-only model explains only $R^2_{\rm RLM}=0.0547$ of the variance in $\log_{10}(M_{\star,\mathrm{IMF}}/M_{\rm dyn,PdA}$). Including velocity dispersion increases the explained variance to $R^2_{\rm RLM}=0.6170$, establishing the $(\log \mathcal{C}, \log \sigma_\star)$ plane as the primary structural relation. The robust best-fitting coefficients are

\begin{equation}
\log_{10}\left(\frac{M_{\star,\mathrm{IMF}}}{M_{\rm dyn,PdA}}\right)    
=
3.149
+
0.173\,\log\mathcal{C}
-
1.389\,\log\sigma_\star.
\end{equation}
also reported in the second line of Table~\ref{tab:regressions}. 

The magnitude of the $\log\sigma_\star$ coefficient is approximately eight times larger than that of $\log\mathcal{C}$, demonstrating that internal kinematics define the dominant axis of variation. At fixed compactness, galaxies with larger velocity dispersions exhibit systematically lower stellar-to-dynamical mass ratios, whereas compactness introduces a comparatively modest geometric modulation.

For completeness, I note that an entirely empirical dynamical-mass proxy constructed directly from $(\log R_{\rm e},\log M_\star)$ and hence without any dependence on $\sigma_{\star}$ (see Appendix~\ref{app:empirical_fit}), yields a qualitatively similar structural ordering, with velocity dispersion remaining the dominant predictor and compactness contributing only a secondary modulation. This confirms that the structural hierarchy is not tied to the specific choice of dynamical-mass prescription. 

\begin{figure}
\includegraphics[width=\columnwidth]{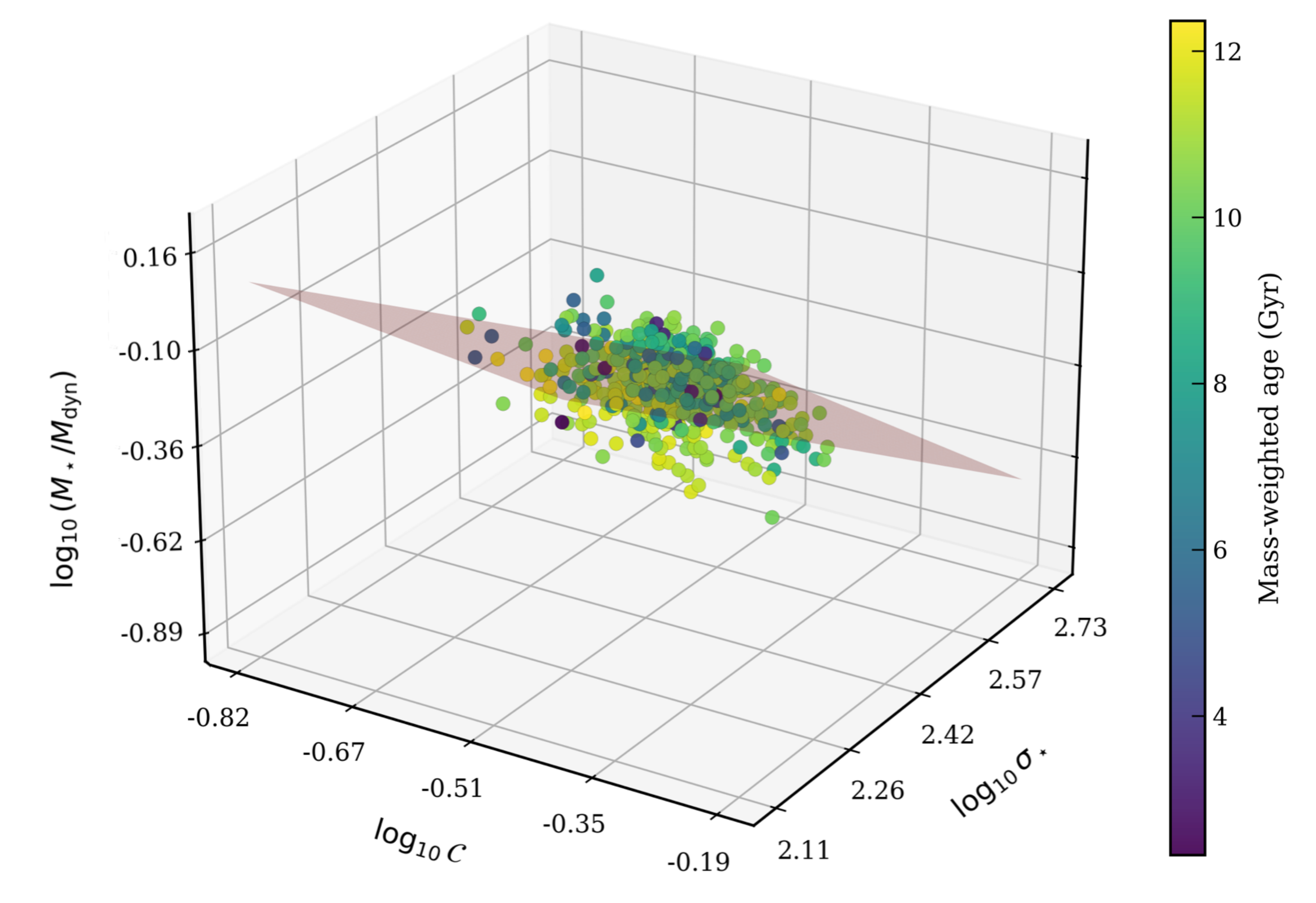}
\caption{
IMF-corrected $\log_{10}(M_{\star}/M_{\rm dyn})$ as a function of
$\log_{10}\mathcal{C}$ and $\log_{10}\sigma_{\star \mathrm{, corr}}$ for the full UCMG sample.
Points are colour-coded by mass-weighted stellar age. 
The semi-transparent surface shows the best-fitting robust plane obtained from Eq.~\ref{equation}.}
\label{fig:hyperplane}
\end{figure}

Figure~\ref{fig:hyperplane} illustrates the resulting structural manifold. The data populate a well-defined tilted surface in $(\log \mathcal{C}, \log \sigma_{\star \mathrm{, corr}})$ space, and the semi-transparent plane corresponds to the best-fitting robust solution of Eq.~\ref{equation}. Once both structural size information and internal kinematics are considered simultaneously, the stellar-to-dynamical mass ratio follows a smooth two-dimensional relation with substantially reduced scatter relative to the compactness-only description.

To assess potential leverage effects, I performed a leave-one-out sensitivity analysis for the structural plane. Removing the galaxies with the largest absolute standardized RLM residuals produces $R^2_{\rm RLM}$ values between 0.608 and 0.622, fully consistent with the full-sample result of 0.6170. The coefficients vary by less than a few percent, confirming that the dominance of $\sigma_\star$ is not driven by a small number of influential systems.

Because $M_{\rm dyn} \propto K\,\sigma_\star^2 R_e$ under virial equilibrium, one may ask whether the strong $\sigma_\star$ dependence arises trivially from algebraic coupling in the mass definition. In Appendix~\ref{app:shuffle}, a Monte Carlo shuffle test demonstrates that the observed coefficient $c_\sigma=-1.389$ lies $\sim 18.9\sigma$ away from the null distribution expected from shuffled $\sigma_\star$ values (median $\simeq -0.0008$, 95\% CI $[-0.147,\,0.137]$), confirming that the $\sigma_\star$ dependence is not a trivial artefact of the virial estimator.

Finally, in Appendix~\ref{app:MC}, I verify that this result is robust to measurement uncertainties in all variables using a MC errors-in-variables treatment. The inferred coefficients remain consistent within 1$\sigma$ of the robust fit, indicating that measurement uncertainties do not materially alter the inferred structural scaling.

\begin{figure}
  \centering
  \includegraphics[width=\columnwidth]{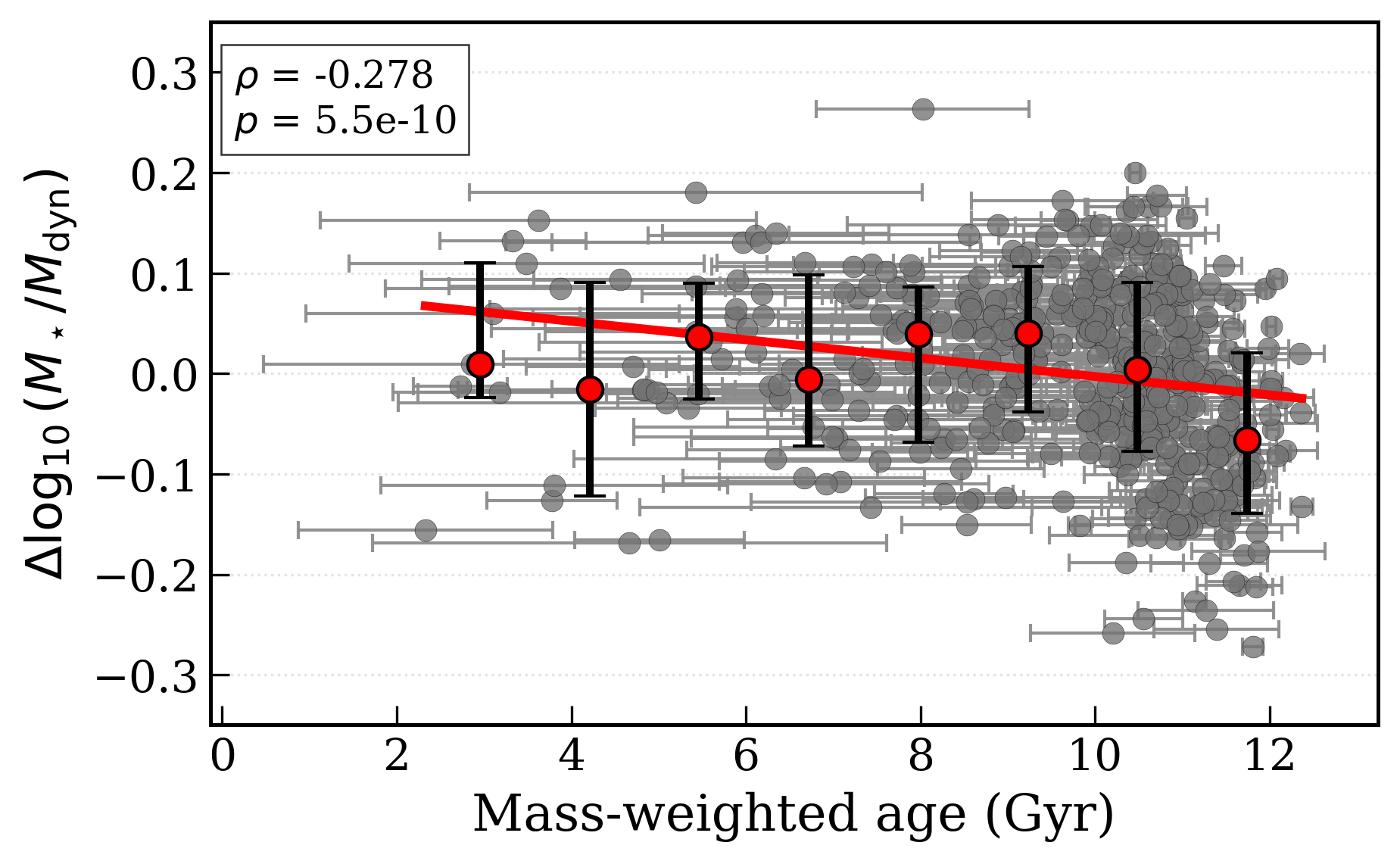}
  \caption{
Residuals of the robust structural plane (Eq.~\ref{equation}) as a function of mass-weighted stellar age. 
Grey points show individual galaxies. Red symbols indicate binned medians with 16--84\% intervals. 
The Spearman coefficient is reported in the top-left corner.
}
  \label{fig:residuals_age}
\end{figure}

Although the structural plane accounts for most of the variance, a mild gradient in stellar age is visible across the surface. Including mass-weighted age increases the explained variance from $R^2_{\rm RLM}=0.6170$ to $R^2_{\rm RLM}=0.6317$, corresponding to $\Delta R^2=0.0147$. The fitted coefficient $d_{\rm Age}=-0.0099$ implies a modest evolutionary tilt, such that older systems exhibit slightly lower $M_\star/M_{\rm dyn}$ at fixed structural parameters. 
Figure~\ref{fig:residuals_age} shows that the residuals from the robust structural plane exhibit a statistically significant but weak anti-correlation with stellar age (Spearman $\rho = -0.278$, $p = 5.5\times10^{-10}$), confirming a small independent modulation.

By contrast, replacing age with the DoR yields no meaningful improvement in explained variance ($\Delta R^2 = 0.0003$), and the fitted coefficient $d_{\rm DoR}$ is statistically consistent with zero. Although age and DoR both trace aspects of assembly history, only mass-weighted age contributes a statistically detectable, albeit modest, independent tilt once compactness and velocity dispersion are included.

In summary, the stellar-to-dynamical mass ratio in UCMGs is primarily governed by internal kinematics. Compactness provides a secondary geometric modulation, stellar age introduces a small evolutionary tilt, and the DoR does not add independent explanatory power beyond these structural and dynamical parameters. The resulting relation is therefore best interpreted as a kinematically anchored structural manifold, with only a modest evolutionary modulation imprinted by stellar age.

\section{Discussion and conclusions}
\label{sec:discussion}

The results presented in this work demonstrate that the stellar-to-dynamical mass ratio in ultra-compact massive galaxies is fundamentally governed by internal kinematics rather than by compactness alone. While compactness correlates with $M_\star/M_{\rm dyn}$ under a homologous virial assumption, this description becomes incomplete once velocity dispersion is included. The emergence of a structural–dynamical manifold in $(\log \mathcal{C}, \log \sigma_\star)$ space indicates that the depth of the gravitational potential, traced by $\sigma_\star$, defines the principal axis of variation in the mass discrepancy.

Under virial equilibrium, $M_{\rm dyn} \propto \sigma_\star^2 R_{\rm e}$ for homologous systems. Deviations from a constant virial coefficient are typically interpreted as structural non-homology. However, the independent role of $\sigma_\star$ in the multivariate regression shows that non-homology in UCMGs cannot be reduced to a pure size effect. Instead, galaxies with similar compactness may differ substantially in their internal dynamical structure, indicating diversity in central mass concentration, orbital anisotropy, or baryonic fraction within the effective radius. Such diversity is consistent with the range of internal orbital configurations observed in compact systems, which can include significant rotational support and anisotropic stellar motions \citep[e.g.][]{Zhu25}. 

The fact that the fitted $\log\sigma_\star$ coefficient is significantly shallower than the analytic expectation associated with pure virial scaling (i.e. a slope near $-2$) is physically meaningful. If the $\sigma_\star$ dependence were purely an algebraic consequence of $M_{\rm dyn}\propto\sigma_\star^2 R_{\rm e}$ under structural homology, one would expect a coefficient close to the analytic baseline. This comparison therefore provides a direct empirical test: if the observed relation were driven purely by the virial definition, the fitted slope should approach the algebraic expectation of $-2$. The significantly shallower coefficient measured here indicates that the $\sigma_\star$ dependence reflects genuine variations in dynamical structure rather than a trivial mathematical coupling. The observed value ($c_\sigma \approx -1.39$) therefore indicates that the effective virial coefficient is not constant across the UCMG population. In realistic galaxies the proportionality between $M_{\rm dyn}$ and $\sigma_\star^2 R_{\rm e}$ depends on the internal mass distribution and orbital structure. Variations in the inner total mass-density slope, the degree of central baryonic concentration (including the contribution of a massive black hole), the dark-matter fraction within $R_{\rm e}$, or the orbital anisotropy can all modify the mapping between the observed line-of-sight dispersion and the total kinetic energy \citep[e.g.,][]{Cappellari+07_SAURONX, Cappellari+13_ATLAS3D_XV, Tortora14, Dutton_Treu14}. As a result, systems with similar compactness can exhibit different dynamical responses to increasing $\sigma_\star$. Disentangling these structural effects requires spatially resolved dynamical modelling capable of separating stellar, black-hole, and dark-matter contributions within $R_{\rm e}$.

UCMGs are not uniformly dynamically hot at fixed stellar mass. Systems identified as relic-like tend to host both older stellar populations and elevated velocity dispersions \citep{Spiniello24}, consistent with deep central potentials established during rapid, dissipative early assembly \citep[e.g.,][]{Naab+09,Oser+10}. Extreme systems such as NGC\,1277, characterised by very high central stellar densities and a dynamically measured supermassive black hole \citep{vandenBosch12,Walsh16}, illustrate how compact baryonic concentrations can significantly enhance measured dispersions at fixed global size. The absence of uniformly high dispersions across the entire UCMG population further supports the conclusion that compactness alone cannot fully describe the dynamical state, and that diversity in internal structure contributes to the observed manifold.

The secondary modulation introduced by stellar age indicates that assembly history leaves a measurable imprint on the structural–dynamical relation. Although modest in amplitude compared to the dominant $\sigma_\star$ dependence, the age trend suggests that earlier-forming systems occupy systematically different regions of the manifold. This behaviour may reflect higher central baryonic fractions or steeper inner mass profiles established during rapid early assembly, as predicted in dissipative formation scenarios \citep{Hopkins+08_DELGN_I, Hopkins+09_DELGN_II, Dekel14, Zolotov15}. Alternatively, residual correlations between IMF normalisation and $\sigma_\star$ \citep{Spiniello+12, LaBarbera+12_SPIDERVII_CG, Spiniello+14} could contribute to the observed age dependence if IMF corrections do not fully remove population-driven mass-to-light variations. Inside-out growth processes may further modify the stellar-to-dynamical balance within fixed apertures, depending on how later accretion redistributes stellar and dark mass in the inner regions. The absence of an independent contribution from the DoR once stellar age is included implies that the relevant evolutionary information is effectively encoded in the integrated stellar age rather than in higher-order burstiness diagnostics. The mass discrepancy therefore reflects coupled variations in dynamical structure and assembly history, with present-day kinematics acting as the primary regulator.

The structural hierarchy identified here is robust against variations in modelling assumptions. The dominance of velocity dispersion is recovered under both a constant virial prescription and the structure-dependent calibration of \citetalias{PeraltaDeArriba+14}, and remains stable whether or not IMF-driven stellar mass corrections are applied. These tests indicate that the structural–dynamical manifold is not an artefact of a particular mass normalisation, but a genuine property of the ultra-compact regime.

Taken together, these findings imply that non-homology in UCMGs encodes dynamical diversity rather than purely geometric compactness. The stellar-to-dynamical mass ratio in the ultra-compact regime is best interpreted as the outcome of coupled structural and assembly processes, with internal kinematics providing the dominant coordinate of variation.

The main results of this study can be summarised as follows:
\begin{itemize}

\item Compactness alone does not fully determine the stellar-to-dynamical mass ratio in ultra-compact massive galaxies. Although size correlates with $M_\star/M_{\rm dyn}$ under a homologous virial assumption, this description becomes incomplete once internal kinematics are taken into account.

\item A multivariate regression in $(\log\mathcal{C},\log\sigma_{\star})$ space accounts for $61.7\%$ of the variance in $\log_{10}(M_{\star,\mathrm{IMF}}/M_{\rm dyn,PdA}$). The scaling is driven primarily by velocity dispersion, indicating that internal kinematics, rather than size alone, regulate the stellar-to-dynamical mass balance in the ultra-compact regime. At fixed compactness, systems with deeper central potentials systematically exhibit lower $M_\star/M_{\rm dyn}$.

\item Stellar age introduces a secondary but coherent evolutionary tilt. Galaxies that assembled earlier tend to occupy regions associated with deeper central potentials, linking present-day dynamical structure to early assembly processes.

\item By contrast, higher-order diagnostics such as the DoR, as well as metallicity, and [Mg/Fe] do not provide independent constraints once size, kinematics, and stellar age are included. The dominant evolutionary information is therefore effectively encoded in the global assembly epoch traced by stellar age.

\end{itemize}

\noindent Taken together, these results define a structural–dynamical–evolutionary manifold in which compactness reflects global size scaling, velocity dispersion anchors the depth of the gravitational potential, and stellar age introduces a modest evolutionary modulation.

Future progress in this field requires spatially resolved dynamical constraints capable of disentangling the relative contributions of stellar mass, dark matter, orbital anisotropy, and central black holes within one effective radius. Given the incredibly small apparent sizes of these objects,  adaptive-optics (AO) assisted integral-field spectroscopy on 8–10\,m class telescopes, and forthcoming facilities such as the ELT, are  essential to resolve the inner kinematic structure of UCMGs at sub-kiloparsec scales. Such data will allow direct measurement of inner mass-density slopes and central mass concentrations, providing decisive tests of whether the observed structural–dynamical manifold reflects variations in baryonic dominance, dark-matter fraction, or black-hole mass fraction.

On the theoretical side, cosmological simulations with sufficient spatial resolution to track dissipative assembly and subsequent structural evolution are required to determine whether the coupled structural–dynamical–evolutionary manifold identified here arises naturally from specific formation pathways or reflects a more universal attractor in the ultra-compact regime.

\begin{acknowledgements}
The author wishes to thank Dr. Crescenzo Tortora, Prof. Michele Cappellari and Prof. Ortwin Gerhard for interesting discussions. 
\end{acknowledgements}

\bibliographystyle{aa} 
\bibliography{biblio_INSPIRE.bib}

\begin{appendix}

\section{Empirical dynamical mass estimator}
\label{app:empirical_fit}

To complement 
the principal analysis based on the structure-dependent virial estimator ($M_{\rm dyn,PdA}$), 
in this Appendix I examine whether the structural–dynamical trends derived in
Sec.~\ref{sec:hyperplane} are preserved when adopting an entirely empirical 
dynamical-mass proxy.  
For this purpose, I  construct a non-virial estimator, $M_{\rm dyn,fit}$, obtained 
from a robust regression between $\log(M_\star/M_{\rm dyn,K=5})$ and 
$(\log R_{\rm e}, \log M_\star)$, and then re-evaluate the structural relations 
using consistently IMF-corrected stellar masses.  
Although this estimator contains no explicit dependence on $\sigma_\star$, it provides 
a useful diagnostic for assessing the stability of the structural manifold under 
changes in the assumed dynamical-mass prescription.

\begin{figure}
  \centering
  \includegraphics[width=\columnwidth]{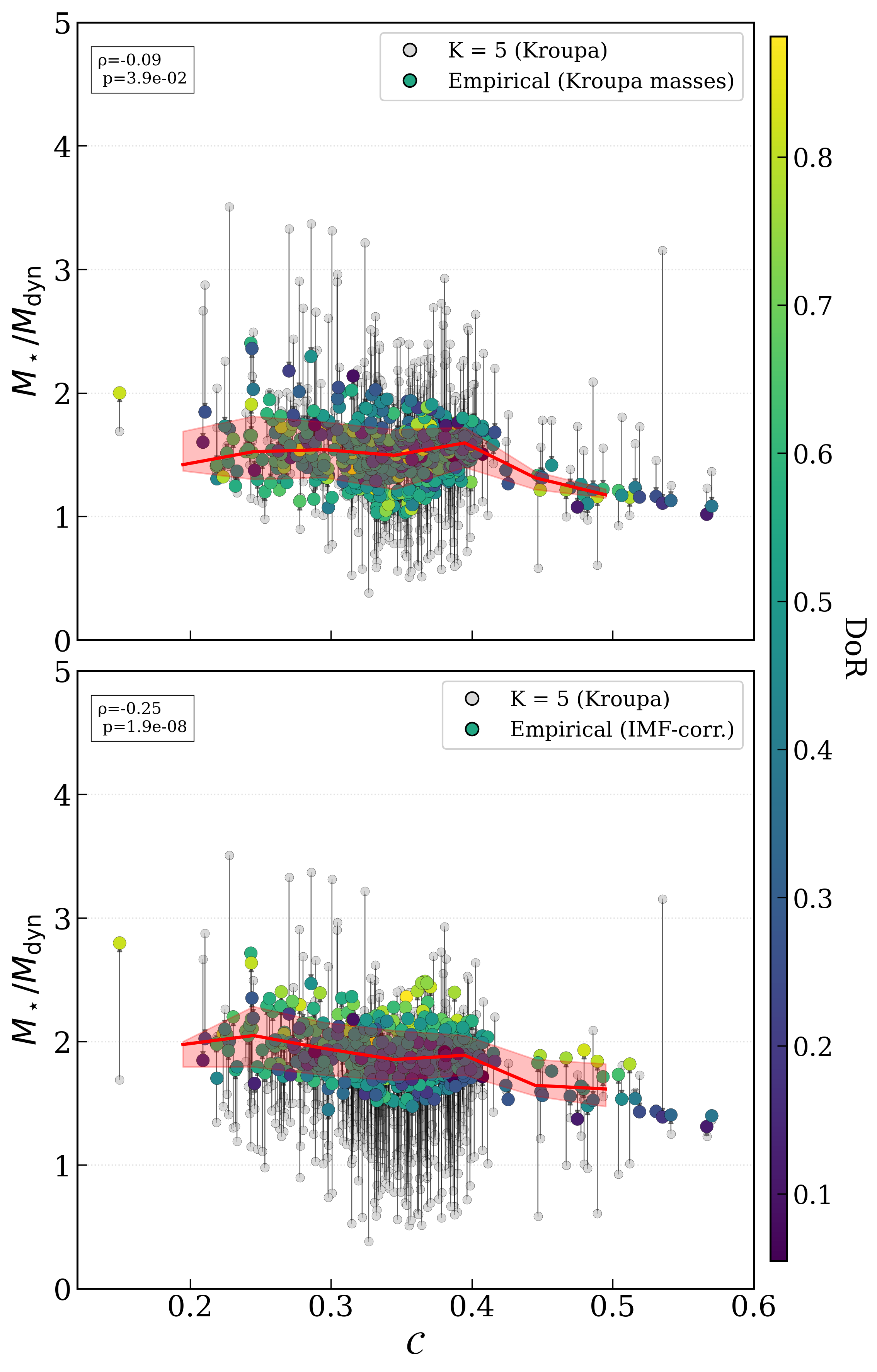}
  \caption{
  Empirical dynamical mass estimator versus compactness. In the top panel the empirical estimator is built using Kroupa IMF for the stellar masses, while in the bottom I apply the correction presented in the main body of the paper. Colour and symbol scheme as in Figure~\ref{fig:compactness}.}
  \label{fig:appa_empirical}
\end{figure}

First, Figure~\ref{fig:appa_empirical} shows the resulting compactness trends when using Kroupa-like IMF (top panel) or IMF-corrected (bot panel) stellar masses. 
As in Figure~\ref{fig:compactness}, the grey points indicate the $K=5$ baseline while coloured points correspond to
$M_{\star}/M_{\rm dyn,fit}$. The thin vertical connectors highlight the object-by-object shift 
between the baseline and the adopted mass definition. 
Fixed-width binned medians and 16--84\% intervals are overplotted in red, and
the full-height colour bar encodes the DoR.

When recomputing the manifold with the new empirical dynamical mass estimator, 
the relation retains a measurable dependence on both compactness and velocity
dispersion, despite the total absence of $\sigma_\star$ in the definition of $M_{\rm dyn,fit}$.  

Fitting the model in Eq.~\ref{equation} to the empirical estimator, 
with a robust Huber–loss regression, I obtain: 
\[
a = -0.590^{+0.062}_{-0.065}, \qquad
b = -0.195^{+0.028}_{-0.026}, \qquad
c = +0.322^{+0.027}_{-0.025},
\]
where the uncertainties are the 16--84\% bootstrap percentiles ($n=1000$).

The empirical-estimator plane explains a non-negligible fraction of the variance
($R^2_{\rm RLM}\simeq 0.378$), although much weaker than the PdA-based relation, as expected from the 
absence of $\sigma_\star$ in $M_{\rm dyn,fit}$. However, and most importantly, the structural hierarchy is 
maintained: 
the unique contribution of velocity dispersion 
$\mathrm{partial}~R^2(\sigma|\mathcal{C})\simeq 0.227$  
exceeds that of compactness  
$\mathrm{partial}~R^2(\mathcal{C}|\sigma)\simeq 0.084$.
Residual correlations with age remain weak
($\rho=-0.078$, $p=0.085$),
while DoR shows a moderate trend 
($\rho=0.223$, $p=7.9\times 10^{-7}$),
consistent with the fact that the empirical estimator does not encode the 
structural non-homology captured by the PdA formalism.

The behaviour of the IMF-corrected estimator in 
Fig.~\ref{fig:appa_empirical} parallels that seen in Figs.~5 and~6:
the compactness dependence, scatter behaviour, and population-residual trends 
remain qualitatively unchanged.  
The structural–dynamical manifold therefore does not rely on the PdA 
dynamical-mass prescription; its qualitative structure persists even when a 
non-virial, data-driven mass proxy is adopted.

\section{Propagation of IMF uncertainties}
\label{app:IMF_MC}
The IMF–DoR calibration adopted in this work was derived by \citet{Maksymowicz-Maciata24} using the so-called ``Golden Sample'', consisting of 39 uncontaminated, old \INSPIRE\ galaxies selected after inspection of index–index diagnostics and removal of objects with contaminated spectral indices (see their Appendix~A and the diamond symbols in their Figures~4 and~6). The resulting fit therefore reflects the IMF behaviour measured in the highest-quality INSPIRE spectra and provides the empirical calibration applied here.

To propagate uncertainties in both the IMF–DoR calibration and the measured DoR values, a Monte Carlo approach was adopted. For each galaxy, $N_{\rm MC}=2000$ realisations were generated. The DoR value was perturbed assuming a Gaussian distribution centred on the measured value, with dispersion equal to the quoted $1\sigma$ uncertainty reported by \citet{Maksymowicz-Maciata24}.

The IMF–DoR relation was treated as
\begin{equation}
\Gamma_b = a\,\mathrm{DoR} + b,
\end{equation}
with coefficients $a$ and $b$ sampled independently from Gaussian distributions centred on their nominal values and with dispersions equal to their reported $1\sigma$ uncertainties.

For each realization, the logarithmic stellar-mass correction was computed as
\begin{equation}
\Delta \log M_\star = k_{\rm IMF} \left(\Gamma_b - \Gamma_{b,\mathrm{Kroupa}}\right),
\end{equation}
where $k_{\rm IMF}$ was likewise sampled within its adopted uncertainty range. A lower bound $\Delta \log M_\star \ge 0$ was imposed to prevent extrapolation towards IMFs lighter than Kroupa.

The corrected stellar mass was then obtained as
\begin{equation}
M_{\star,\mathrm{IMF}} = M_\star \times 10^{\Delta \log M_\star}.
\end{equation}

For all analyses in the main text, the IMF-corrected stellar mass was taken as the median of the Monte Carlo distribution, with uncertainties quoted from the 16th–84th percentile range. Stellar-to-dynamical mass ratios were evaluated for each realization, ensuring consistent propagation of IMF-driven uncertainties throughout the analysis.

\section{Robustness tests of the structural–dynamical manifold}
The structural–dynamical manifold identified in Section~\ref{sec:hyperplane} rests on the dominant role of the $\sigma_\star$ coefficient in the multivariate regression. To ensure that this result is not driven by formal properties of the virial definition or by measurement uncertainties, two complementary robustness tests were performed. The first addresses potential algebraic coupling inherent in the dynamical mass estimator, while the second evaluates the impact of propagated observational errors.
\subsection{Robustness of the $\sigma_\star$ dependence to algebraic coupling}
\label{app:shuffle}

Because dynamical mass estimates satisfy $M_{\rm dyn}\propto K\sigma_\star^2 R_e$, algebraic covariance can in principle induce a dependence of $\log(M_\star/M_{\rm dyn})$ on $\log\sigma_\star$. To ensure that the dominant $\sigma_\star$ coefficient measured in Section~\ref{sec:hyperplane} does not arise trivially from this identity, two complementary tests were performed.

First, under the null hypothesis that stellar mass is uncorrelated with $(\sigma_\star, R_e)$, the expected slope of $\log(M_\star/M_{\rm dyn})$ versus $\log\sigma_\star$ is approximately $-2$ (modulo the exact sample distribution and the adopted virial coefficient).

Second, a Monte Carlo shuffle test was conducted. The IMF-corrected stellar masses were randomly permuted among galaxies while preserving the marginal distributions of $M_\star$, $\sigma_\star$, and $R_e$. For each realization, the Peralta de Arriba (PdA) virial coefficient $K_{\rm PdA}(R_e, M_\star)$ was recomputed, $M_{\rm dyn}$ and the ratio $M_\star/M_{\rm dyn}$ were recalculated, and the same robust multivariate model used in the main text was refitted. Results are based on $N=1000$ realisations (seed fixed).

The null distribution of the fitted $\sigma_\star$ coefficient has mean $\langle c_\sigma\rangle=-2.09$ and standard deviation $\sigma_{\rm null}\simeq0.08$. The observed coefficient, $c_\sigma^{\rm obs}=-1.389\pm0.056$, lies $\simeq9\sigma$ away from the shuffle mean. The fraction of realisations producing a coefficient as shallow as the observed value is effectively zero (empirical upper bound $p<1/(N+1)\lesssim10^{-3}$ for $N=1000$). The dominance of $\sigma_\star$ reported in Sect.~\ref{sec:hyperplane} is therefore not an algebraic artefact of the virial estimator, but reflects genuine structural–dynamical coupling.

\subsection{Robustness of the structural plane to measurement uncertainties}
\label{app:MC}

In Sect.~\ref{sec:hyperplane}, the structural manifold relating the stellar-to-dynamical mass ratio to compactness and velocity dispersion was fitted using a robust Huber-loss estimator. Because all three quantities are derived from measured observables and carry non-negligible uncertainties, the inferred regression coefficients could in principle be affected by errors-in-variables (EIV) bias.

In particular, both $C$ and $M_{\rm dyn}$ depend on the effective radius $R_{\rm e}$, while the virial estimator satisfies $M_{\rm dyn}\propto\sigma_\star^2 R_{\rm e}$, introducing covariance between predictors and response. Although Appendix~\ref{app:shuffle} demonstrates that the dominant $\sigma_\star$ dependence is not a trivial algebraic consequence of the virial definition, it remains necessary to verify that measurement uncertainties do not materially alter the inferred slopes.

To assess this, a Monte Carlo EIV analysis was performed. For each galaxy, the observed quantities $\log C$, $\log \sigma_\star$, and $\log_{10}(M_\star/M_{\rm dyn})$ were perturbed assuming Gaussian measurement uncertainties. Uncertainties on $\log \sigma_\star$ were propagated as

\begin{equation}
\delta \log \sigma_\star =
\frac{\delta \sigma_\star}{\sigma_\star \ln 10}.
\end{equation}

Uncertainties on $\log_{10}(M_\star/M_{\rm dyn})$ were obtained by combining in quadrature the IMF-driven stellar-mass uncertainties and the propagated dynamical-mass uncertainty, dominated by the $\sigma_\star^2$ term. Compactness uncertainties were propagated from the effective-radius measurements.

A total of $N=4000$ realisations were generated for the full sample ($N_{\rm gal}=482$). In each realization, perturbed values were drawn and the linear model

\begin{equation}
\log_{10}\left(\frac{M_\star}{M_{\rm dyn}}\right)
=
a + b\,\log C + c\,\log \sigma_\star
\end{equation}

was refitted using ordinary least squares.

The median coefficients and 68\% intervals are
$a = 2.56 \pm 0.13$, 
$b = 0.21 \pm 0.05$, and 
$c = -1.14 \pm 0.06$.
For comparison, the robust RLM fit reported in Section~\ref{sec:hyperplane} yielded
$a = 3.15$, $b = 0.17$, and $c = -1.39$.

Accounting for measurement uncertainties modestly reduces the absolute value of the $\log \sigma_\star$ coefficient and increases its uncertainty, as expected from regression dilution. However, the qualitative structure of the relation remains unchanged: the $\sigma_\star$ term continues to dominate over compactness in absolute magnitude, and its 68\% interval remains well separated from zero.

In all realisations, $|c| > |b|$, confirming that internal kinematics remains the primary axis of variation even after propagating observational uncertainties. The structural–dynamical manifold is therefore robust to measurement-error effects.

\end{appendix}
\end{document}